\documentclass[
  twoside,
  twocolumn,
  aps,
  preprintnumbers,
 floatfix,
 showkeys,
]{revtex4}

\usepackage[dvips]{graphicx}
\usepackage{dcolumn}
\usepackage{color}
\usepackage{amsmath}

\usepackage{tabularx}
\usepackage{lipsum}
\usepackage{rotating}

\usepackage[version=3]{mhchem}

\begin{document}

\title{EXAFS Experiments on Local Structure of NO Adsorbed on Supported Silver Clusters}
\author{Amgalanbaatar Baldansuren$^{\dag}$}
\email[e-mail:\quad]{amgalanbaatar.baldansuren@manchester.ac.uk}
\altaffiliation{$^\dag$Previous address:\quad Institut f\"ur Physikalische Chemie, Universit\"at Stuttgart, D-70569 Stuttgart, Germany}
\affiliation{The Photon Science Institute, EPSRC National EPR Facility and Service, School of Chemistry, The University of Manchester, Oxford Road, Manchester M13 9PL, UK}
\date{\today}

\begin{abstract}
{\it In-situ} EXAFS experiments were performed at the Ag $K$-edge after adsorbing \ce{NO} on the hydrogen reduced silver cluster. The coordination parameters, i.e. $R$({\AA}), $N$ and $\Delta \sigma ^2$({\AA}$^2$), were derived from a three-shell fitting model including a split Ag-Ag shell and a single Ag-O(N) shell. The multiple-shell fitting proved that the reduced silver cluster underwent a slight change in the local structure. After co-adsorbing \ce{NO} and \ce{O2}, the difference file fitting technique was used for isolating each of the first Ag-Ag(1) and Ag-O(N) shell contributions to the EXAFS spectra filtered in $R$-space of Fourier transform.
\end{abstract}

\maketitle

\section{Introduction}

Novel optical, electronic, magnetic and conductive properties appear in materials when structures are formed with sizes comparable to the nanometer scale, dominated by quantum confinement of the electron wavefunction.

The paramagnetic silver clusters with unpaired electrons were expected to be chemically and catalytically active \cite{Tka08,Amg09,Bal09,Bam09,Bal15}. Among others, the paramagnetic six-atom cluster exhibited more activities against nitrogen monoxide than other silver clusters \cite{Bal09}. Our electron paramagnetic resonance (EPR) studies suggested that the adsorbed NO is on a previously diamagnetic and therefore a magnetically silent silver cluster. Furthermore, electron spin echo envelope modulation (ESEEM) methods allowed us to identify the small hyperfine couplings, of the order of nuclear Zeeman and nuclear quadrupole interactions, of $^{14}$N nuclei that were distributed even more farther away from the unpaired electron of the six-atom silver cluster. These results suggested that the reduced coordination of low-nuclearity paramagnetic clusters is an important characteristic to maintain the open-shell configuration, correlated well with the \emph{electronic size effect}.

The \emph{surface size effect} arises from the boundary condition for the confined electronic wavefunction at the surface. For sufficiently small metal clusters the relative merit is a fraction of atoms at the surface, and the low mean coordination number in turn implies a large fraction of the surface atoms \cite{Rod06}. This surface characteristic features uniquely and is seen neither in ordinary bulk metals nor in compounds containing many complicated electronic bands. The properties of the surface atoms basically depend on the type of supports in particular. Micro-porous zeolite supports have the advantages of preserving the metal cluster structure with the controlled size and distribution which are one of the most important factors in an activity and selectivity of heterogeneous catalysts \cite{Cor08}.

Extended X-ray absorption fine structure (EXAFS) is a part of X-ray absorption spectroscopy (XAS) and provides information about electronic and local structure of gas-adsorbed metal cluster catalysts under various conditions \cite{Oud00}. More specifically, the coordination number and inter-atomic distance are the main structural parameters used extensively to evaluate the mean size and surface construction of small metal clusters. This technique can detect up to five nearest coordination shells surrounding absorber atoms \cite{Jen99}. Since EXAFS is an averaging technique, it detects an ensemble of Ag clusters with the nearly same atomic size and structure, no matter whether these clusters are diamagnetic or paramagnetic according to their electron spin configuration and unpaired density distribution of silver clusters. The high sensitivity of EXAFS experiments provided us with accurate information of the local structure and coordination, which were significantly altered following \ce{NO} adsorption \cite{Bam09}. High nitric oxide (NO) adsorption is of great interest for environmental applications in gas separation and NO$_x$ traps for lean burn engines. NO is also an extremely important agent in biology, e.g. in the cardiovascular, nervous, and immune systems \cite{Mon91}, and a storage of NO in solids has potential applications as antithrombosis materials \cite{Kee03}.

\section{Experimental}

The NaA (Si/Al = 1) zeolite was supplied by CU Chemie Uetikon AG in Switzerland. Zeolite samples were heated up in air at a rate of 0.5 K min$^{-1}$ to 773 K where they were kept for 14 hours in order to burn off any organic impurities. Subsequently, 7 g of the heated sample was washed by stirring in 150 ml bi-distilled water containing 40 ml \ce{NaCl} (10\%) solution and 2.76 g of \ce{Na2S2O3\cdot5H2O} salt at 343 K. This washing processes was repeated at least nine times. The washed sample was dried in air at 353 K for 24 hours.

Ag/NaA samples were prepared in a flask containing 2.25 g of pre-treated zeolite by aqueous ion-exchange with 50 ml 50 mM \ce{AgNO3} solution (ChemPur GmbH in Germany, 99.998\%) by stirring at 343 K in the dark for 24 hours. The ion-exchanged sample was filtered and rinsed with deionized water several times, and dried in air at 353 K overnight. Chemical analysis by atomic absorption spectroscopy (AAS) demonstrated that the ion-exchange reaction leads to a silver loading of ca. 12\% (wt.). Oxidation was performed under a gas stream of \ce{O2} (Westfalen AG in Germany, 99.999\%) with a flow rate of 17 ml min$^{-1}$ g$^{-1}$ from room temperature up to 673 K using a heating rate of 1.25 K min$^{-1}$ where it was kept for an additional hour. While the sample was held at the final temperature, the residual \ce{O2} gas in the reactor was purged by \ce{N2} (Westfalen AG, 99.999\%) gas for 1 hour. At 673 K, the heat treatment under oxygen leads to the extensive color changes from white to reddish yellow or dark yellow of Ag/NaA and it is a strong indication of the silver cluster formation. More importantly, XRD measurements proved the NaA zeolite structure remained intact even after such a high temperature treatments (diffraction pattern not shown).

The XAS measurements were performed at the Ag K-edge (25.514 keV) at the Swiss-Norwegian Beamline (SNBL) BM01B of the Storage Ring at the European Synchrotron Radiation Facility (Grenoble, France). The X-rays were passed through a Si(111) channel-cut monochromator and a chromium-coated mirror to reject the higher harmonics. All measurements were performed in transmission mode using ionization chambers for detection. A spectrum of a silver foil was acquired simultaneously with each measurement for energy calibration using a third ionization chamber.

The oxidized catalyst samples of 12\% (wt.) Ag/NaA were pressed into a self-supported wafer form with optimal thickness and positioned in a {\it in-situ} cell \cite{Kam89}. All XAS spectra of oxidized silver clusters were collected under dynamic vacuum (or simply evacuation) at 298 K. The sample was further reduced under an atmosphere of static hydrogen (5\% \ce{H2} in \ce{He}), and XAS spectra were collected under 500 mbar partial pressure. The reduced sample was exposed to 1\% \ce{NO} in \ce{He}, and the spectra were collected under 1000 mbar partial pressure at 298 K. Subsequently, oxygen (5\% \ce{O2} in \ce{He}) was exposed to the nitrogen oxide adsorbed sample. Complying with the safety regulations, all of diluted gases were supplied by the Storage Ring.

XAS data analysis was carried out using the commercially available XDAP software \cite{Var95}. The adsorption data were background-subtracted by means of standard procedures \cite{Kon00}. The spectra were normalized on the height of the edge step at 100 eV above the edge. Multiple-shell fitting was performed in $R$ space ($1.5 < R < 4.0$ {\AA}) using a $k$-weighting of 1, 2, and 3 and a $k$ range of $3 < k < 14$ {\AA}$^{-1}$ \cite{Tro02}. A silver foil and \ce{Ag2O} were measured as references and analyzed using the FEFF8 \cite{Ank98}.

\section{Results and discussion}

The only results of EXAFS at the Ag $K$-edge of the oxidized and reduced six-atom clusters were reported in our peer-reviewed article \cite{Bam09}. The highest fraction of the surface atoms implies a low mean coordination number of $N\approx4.0$, which is accompanied by a direct consequence of the bond length contraction of Ag-Ag in the cluster. In total, the distance of the split Ag-Ag shell is contracted by $\sim5-7\%$ as compared to the bulk distance of 2.889 {\AA}. This effect is generally caused by the increased electron density between the atoms because of the re-hybridization of the $s$, $p$, and $d$ metal orbitals \cite{Del83}.

\begin{figure}[ht]
\centering
\includegraphics[width=0.6\columnwidth]{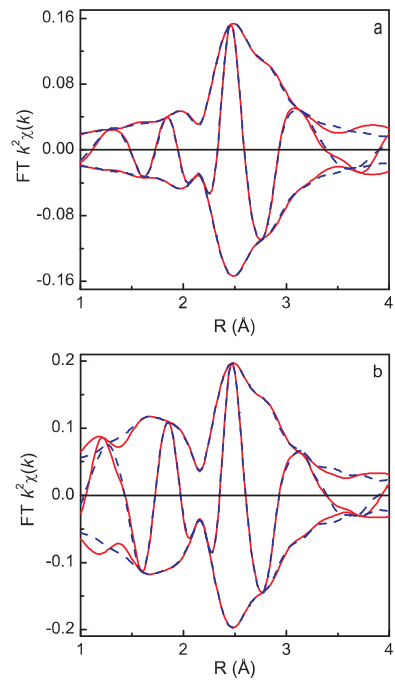}
\caption{Comparison between the FT of the $k^2$-weighted $\chi(k)$ function ($k^2$, $\Delta k$ = 3.0 - 14 {\AA}$^{-1}$) of the experimental EXAFS (solid line) with their theoretical fits (dashed line) with a split Ag-Ag shell and a single Ag-O(N) shell. a) after adsorbing \ce{NO} on the cluster at 298 K. b) after exposing the \ce{NO} adduct cluster to \ce{O2}.}
\label{fig:XAS1}
\end{figure}

As N and O atoms are consecutively located in the periodic table, their backscattering amplitude and phase shifts are expected to be very similar. In the fit, a contribution of nitrogen and oxygen is represented as a single Ag-O(N) shell (Table \ref{tab:XAS}). After adsorbing \ce{NO} in \emph{ex-situ}, contributions of these atoms were identifiable from our pulsed EPR experiments since oxygen is not magnetic with a zero nuclear spin \cite{Bal09}. The intense peak at 2.5 {\AA} in $R$-space of Fourier transform of the $k^2$-weighted $\chi(k)$ function represents the split Ag-Ag shell, while the Ag-O(N) shell is also observable  at 1.9 {\AA} (Figure \ref{fig:XAS1}a). The coordination of the second Ag-Ag shell is of the order of $N\approx3.2$, indicating that only $<0.2$ atom was displaced from its position in this shell \cite{Bam09}. The coordination of the first Ag-Ag shell remained intact with $N\approx0.40$. The inter-atomic distances were almost the same 2.77 {\AA} and 2.69 {\AA} as previous for both shells. The local structure withstands the adsorption reaction, and the silver cluster exhibits the same atomic size as the reduced cluster. EXAFS is an averaging technique, and its detection is not limited as to whether the NO adduct is adsorbed on diamagnetic or on paramagnetic clusters. It is therefore concluded that all six Ag atoms are still at the cluster surface, thus possessing the same coordination numbers at virtually unchanged inter-atomic distances of the split Ag-Ag shell \cite{Bam09}. A higher Debye-Waller factor corresponds to a large static disorder due to a high fraction of Ag atoms at the cluster surface. The coordination of the Ag-O shell plus nitrogen is of the order of $N\approx0.80$ at a distance of 2.29 {\AA}. This distance reflects the bond length between silver and nitrogen atoms (Ag$-$N), compatible with the value obtained from the point-dipole approximation for the electron and nuclear spin magnetic dipoles \cite{Bal09}. The value is considerably smaller than the oxygen shell coordination of the reduced silver cluster \cite{Bam09}. This indicates that the silver clusters were just partially reduced by \ce{H2} and became more reduced in the local structure when adsorbing \ce{NO}. This also suggests that there was a number of oxygen atoms interspersed between the silver atoms of the cluster and decreased by adding more reductive gas molecule. Above 3.5 {\AA}, a negligibly small deviation between the theoretical and experimental spectra is actually due to a contribution of light scattering N and O atoms (Figure \ref{fig:XAS1}a). Nevertheless, the fits are very good, especially the imaginary parts. The imaginary part is used to judge the fit quality \cite{Var95}, as long as this is sensitive to the inter-atomic distance.

\begin{figure}[ht]
\centering
\includegraphics[width=0.6\columnwidth]{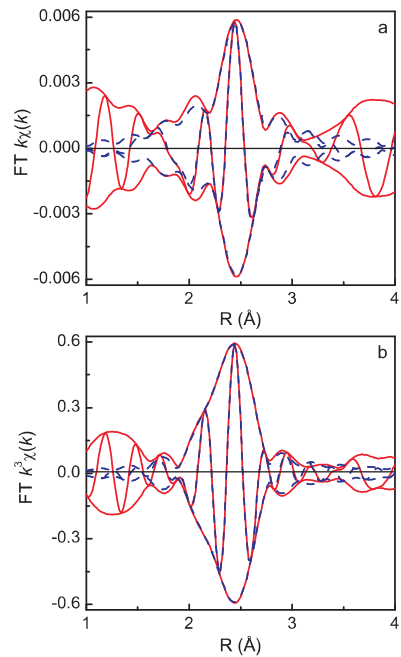}
\caption{EXAFS difference file spectra of the gas adsorbed clusters. Fourier transform of a) $k^1$- and b) $k^3$-weighted $\chi(k)$ functions ($\Delta k$ = 3.0 - 14 {\AA}$^{-1}$, no corrections for phase and amplitude of Ag-Ag). Raw data minus fitted Ag-Ag(2) and Ag-O(N) contributions (red lines) and the best fits of Ag-Ag(1) (blue lines).}
\label{fig:XAS2}
\end{figure}

By exposing the NO adsorbed cluster to \ce{O2}, the silver cluster structure experienced further changes in the local structure (Figure \ref{fig:XAS1}b). The same effect was observed that the EPR signal of the NO adduct shifted to the completely new hyperfine spectrum of the active species, namely paramagnetic \ce{NO2} after exposing that sample to \ce{O2} at room temperature \cite{Bal09}. The coordination of the first shell remained unchanged at the same distance of 2.70 {\AA}. In the second Ag-Ag shell, the silver coordination increased to $N\approx4.50$ at an average distance of 2.76 {\AA}, indicating that the silver cluster became less compact. Thus, the total Ag-Ag coordination increased to a certain extent. A major change occurred to the Ag-O(N) shell as its coordination also increased to $N\approx2.10$ at an average distance of 2.26 {\AA}. The intensity of this shell increased significantly in the $R$-space spectrum, suggesting that the silver clusters were oxidized again upon oxygen exposure. More importantly, the coordination difference before adding oxygen demonstrates again the silver to nitrogen (Ag-N) coordination. The contribution of Ag-O(N) shell probably arose from only N atom, and hence one NO molecule was adsorbed per cluster.

\begin{table*}[t]
  \centering
  \caption{Structural parameters obtained from the three-shell fittings for oscillatory $\chi(k)$ function before and after exposing the reduced Ag clusters in 12 wt\% Ag/NaA to adsorbate molecules at room temperature.}

  \label{tab:XAS}
  \begin{tabular}{*{10}{lccccccccc}}
    \hline
                        &          & Ag-Ag  &                       &          & Ag-O(N)  &                         &          & Ag-Ag   &                            \\
    Samples             & $R$      & $N$    & $\Delta\sigma^2$      & $R$      & $N$      & $\Delta\sigma^2$        & $R$      & $N$     & $\Delta\sigma^2$           \\
                        & (\AA)    &        & (\AA$^2$)             & (\AA)    &          & (\AA$^2$)               & (\AA)    &         & (\AA$^2$)                  \\
    \hline
    NO adsorbed         & 2.77     & 3.2(3) &0.012                  & 2.29     & 0.8(2)   & 0.009                   & 2.69     & 0.4(3)  & 0.003                       \\
    NO+O$_2$ adsorbed   & 2.76     & 4.5(5) &0.011                  & 2.26     & 2.2(3)   & 0.013                   & 2.70     & 0.3(3)  & 0.007                       \\
    \hline
    $R$(\AA) - bond length                           \\
    $N$ - coordinations                              \\
    $\Delta\sigma^2$(\AA $^2$) - Debye-Waller factor \\
  \end{tabular}
\end{table*}

In order to show the contribution of oxygen plus nitrogen, the difference file technique was used \cite{Kon00}. The contribution of Ag-Ag is more clearly seen from the difference file spectrum shown in Figure \ref{fig:XAS2}b. The phase and amplitude correction were not applied since the quality of the raw EXAFS data was good enough. This is a good fit that only a slight asymmetry is observed in Fourier transform of the $k^3$-weighted $\chi(k$) function spectrum in the range of $\Delta k$ = 3.0 - 14 {\AA}$^{-1}$. The real and imaginary parts are fitted reasonably good, showing the Ag-Ag(1) shell contribution. There was an increase in the coordination of the second Ag-Ag shell, indicating that added oxygens resulted in a re-dispersion of silver clusters located in neighboring sites. It indicates that the cluster environment becomes more uniform as zeolite contains oxygens in the frameworks. In some cases the high correlation between $N$ and $\Delta \sigma^2$ makes difficulties in getting a good fit with a unique set of coordination parameters for the local structure. Such a unique set of parameters can be harmonized and checked using the different $k$-weighting factor for the difference file spectra. In the $k^1$-weighted spectrum, there is a significant residue of the light scatterer contribution such as N and O (Figure \ref{fig:XAS2}a). There is a mismatch between the theoretical and experimental spectra of the Ag-O(N) contribution. This is acceptable since the NO adduct shows molecular dynamics, which affects the fit quality of the difference file spectrum. Besides, the first Ag-Ag shell has less contact with adsorbates. XANES spectra demonstrated that the absorption edge position and the white-line intensity were not, or only slightly changed during the {\it in situ} adsorption, indicating that the clusters remained reduced (spectrum not shown).

\section{Conclusion}

The coordination number of a split Ag-Ag shell provides the mean size of silver clusters. By using the multiple scattering path to analyze the absorption fine structure data, the coordination parameters for the local structure of silver clusters were determined with high accuracy. The reduced clusters are quite mono-disperse consisting of $6\pm1$ Ag atoms, while all atoms belong to the cluster surface.

Adsorption of \ce{NO} on the reduced silver cluster is discussed for the first time by utilizing {\it in-situ} EXAFS measurements. The silver cluster underwent a slight change in the local structure, and the Fourier transform of the $k^2$-weighted $\chi(k)$ function clearly showed the contribution of the Ag-O(N) and Ag-Ag shell. Upon NO adsorption, the silver cluster also became more reduced as for the local structure because the oxygen remained partly interspersed between the silver atoms after hydrogen reduction.

In order to demonstrate the presence of both the heavy and light scattering atoms, the difference file technique was used for isolating and fitting the Ag-O(N) and the Ag-Ag(1) shell contribution to the spectrum in $R$-space of Fourier transform after co-adsorbing \ce{NO} and \ce{O2}. This isolating technique also provides a reliable set of coordination parameters by applying the different $k$-weighting factor. The coordination parameters of the cluster led to observable changes in the local structure but not the drastic ones, indicating that the cluster rearrange up to one or two silver atoms in the second shell.

\section*{Acknowledgement}

The author is thankful to Prof. Emil Roduner for his valuable discussions. The author is grateful to Prof. J. van Bokhoven for his valuable discussions on X-ray absorption spectroscopy results. A.B. was supported by the Deutsche Forschungsgemeinschaft for a doctoral thesis scholarship through the Research Training Group 448 "Advanced Magnetic Resonance Type Methods in Materials Science" at the University of Stuttgart.

\end{document}